\documentclass[twocolumn,showpacs,preprintnumbers,amsmath,amssymb,superscriptaddress]{revtex4}

\usepackage{graphicx}

\usepackage{dcolumn}
\usepackage{bm}
\usepackage{amsmath}
\usepackage{color}

\begin{document}

\author{William N. Plick}

\affiliation{Quantum Optics, Quantum Nanophysics, Quantum Information, University of Vienna, Boltzmanngasse 5, Vienna A-1090, Austria}
\affiliation{Institute for Quantum Optics and Quantum Information, Boltzmanngasse 3, Vienna A-1090, Austria}

\author{Radek Lapkiewicz}

\affiliation{Quantum Optics, Quantum Nanophysics, Quantum Information, University of Vienna, Boltzmanngasse 5, Vienna A-1090, Austria}
\affiliation{Institute for Quantum Optics and Quantum Information, Boltzmanngasse 3, Vienna A-1090, Austria}

\title{An explicit contextualized hidden variable model replicating an indivisible quantum system}

\begin{abstract}
\noindent Recent experiments and theory have further illuminated the concept of ``quantum contextuality". In this paper we take an inequality \--- the Pentagram (or KCBS) inequality, which is violated by an unentangled spin-1 system \--- and given a relaxed assumption of non-contextuality show that a hidden variable model may be constructed that replicates exactly the quantum bound. This is in contrast to the case of an entangled pair of spin-1/2 particles where a similar relaxation of the assumption of locality does not lead to a replication of the quantum bound. Some reasons are proposed for why this may be the case.
\end{abstract}


\maketitle

\section{Introduction}

\noindent With a seminal paper in 2008 Klyachko, Can, Binicioglu, and Shumovsky (KCBS) showed how an indivisible (single-particle, unentangled) spin-1 system could be used to experimentally show contradiction with a fundamental assumption of the classical world-view: namely that of non-contextual realism; via an inequality \cite{pent}. Two recent experiments have shown such a violation \cite{exp1,exp2}, proving quantum contextuality in these systems.\\

\noindent The question of \emph{how much} contextuality is required to violate this inequality now naturally arises. As such it is useful to first reiterate briefly, what exactly contextuality means. An outcome of a particular measurement may be considered to be contextual if it depends on a separate outcome and/or measurement. For example, take the canonical case of an entangled pair of spin-1/2 particles. The assumption of local-realism states that, once the measurement events are space-like separated, a measurement performed on the first particle can not affect the outcome of a measurement on the second particle. This is a form of non-contextuality, specifically \--- that the second measurement ($b$) and outcome ($B$) has no contextual relationship with the measurement ($a$) and outcome ($A$) of the first. Thus, locality is a subclass of non-contextuality, reinforced by a physical motivation. The assumption of non-contextual realism leads to the famous Bell \cite{bell} and CHSH \cite{chsh} inequalities.\\

\noindent Apart from the assumption of locality there are other forms of non-contextuality. For the case of a spin-1 particle, non-contextuality may be defined in terms of what observables are co-measurable (i.e. which observables commute). That is, if $[\hat{A},\hat{B}]=0$, $[\hat{A},\hat{C}]=0$, but $[\hat{B},\hat{C}]\neq 0$ then we say that a measurement of $\hat{A}$ is co-measurable with both $\hat{B}$ and $\hat{C}$ (though $\hat{B}$ and $\hat{C}$ are not). Thus, measurements of $\hat{A}$ are non-contextual with $\hat{B}$ (or $\hat{C}$). For a more in-depth discussion see, for example, Ref.\cite{peres}.\\

\noindent Experimentally, a controllable spin-1 system may be implemented, for example, by splitting a photon between three spatial modes. The set of commuting operators then simply become heralded detections on these modes. For more details see Ref.\cite{exp1}.\\

\noindent However it could be argued that this assumption of non-contextuality via co-measurability is philosophically weaker than the assumption of locality since the locality assumption has a very strong physical basis. Thus there is some motivation for either finding a more physically robust assumption on which to base the construction of the inequality, or manipulating the physical implementation of the spin-1 system such that the assumption of non-contextuality is transmuted to an assumption of locality. The latter is possible by space-like separating the measurement events on the photonic modes, but this is not the subject of this paper.\\

\noindent For an entangled pair of spin-1/2 particles it is possible to construct an explicit non-local hidden variable model which replicates the quantum bound of the CHSH inequality. It is also possible to show that an inequality obeyed by this specific model (as well as all others of the same class) is in contradiction with quantum physics. The derivation of this inequality \--- the Leggett inequality \--- proceeds by requiring that the individual photons obey the spin projection rule on the marginal probabilities, which for a photonic spin-1/2 system is the well known Malus law for polarization \cite{leg}. Alternatively the derivation may proceed by assuming that marginal probabilities may not take negative values \cite{bra}.\\

\noindent It was the original goal of the research reported here to find a similar inequality for the photonic spin-1 system. Where the assumption of non-contextuality is replaced with an assumption with a clear physical motivation: that of obedience of the spin-projection rules, a physically provable property. Instead what was found was that such a procedure can not succeed for the pentagram inequality, demonstrating interesting differences between the spin-1 and entangled spin-1/2 systems. This, however, does not rule out the possibility that differently structured inequalities (perhaps with more measurement contexts \--- though more measurement \emph{directions} will not be useful, as we shall see) could discriminate between semi-contextual realism and quantum mechanics.\\

\noindent In the following section we review the derivation of the pentagram inequality with emphasis on the photonic implementation. In section 3 we derive a contextual inequality (restricted as in the Leggett inequality to following the spin-projection rules on the marginal probabilities) for a spin-1 system which replicates the quantum bound (i.e. not violated by QM) \--- in contrast to the Leggett inequality, which \emph{is} violated by quantum mechanics. In section 4 we derive an explicit hidden variable model which replicates exactly quantum physics for the system in question. In section 5 we review our results and offer some potential explanations for the behavior of this model.

\section{The Pentagram Inequality}

\noindent Consider a single spin-1 particle. The operators representing spin-squared measurements along three orthogonal real directions (e.g. $\hat{S}_{x}^{2}$, $\hat{S}_{y}^{2}$, $\hat{S}_{z}^{2}$) commute and are thus co-measurable. These operators act on states which may be represented as vectors in $\mathbb{C}^{3}$. Similarly we may take three co-measurable projection operators working on a single photon split between three spatially separate optical modes (eg. $|x\rangle\langle x|$, $|y\rangle\langle y|$, $|z\rangle\langle z|$). In the latter case we may also picture the operators as measurements along real directions. \\

\noindent Now take operators of the form $\hat{a}_{j}\equiv2|j\rangle\langle j|-1$, where $j$ is some direction in real-space. We have $[\hat{a}_{j}, \hat{a}_{k}]=0$ if $j$ and $k$ are orthogonal directions, thus the $a$'s are co-measurable for orthogonal directions. A single measurement of $\hat{a}_{j}$ will yield +1 or -1, depending on whether there is, or is not, a photon in the optical mode represented by the direction $j$. Non-contextually, we could make a series of five measurements

\begin{eqnarray}
a_{1}a_{2}+a_{2}a_{3}+a_{3}a_{4}+a_{4}a_{5}+a_{5}a_{1}.
\end{eqnarray}

\begin{center}
\begin{figure}[h]\centering
\includegraphics[scale=0.5]{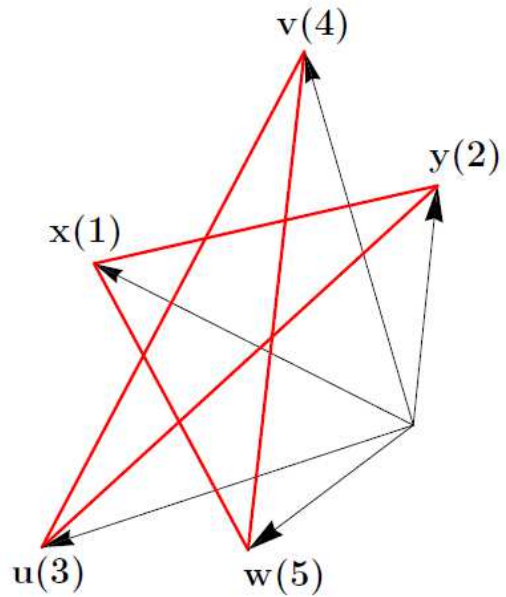}
\caption{\label{pent1}(Color online) Five pairwise-orthogonal directions visualized as a pentagram. Each direction is orthogonal to the two directions connected to it by the pentagram. The directions themselves are unit vectors labeled by by letters (where $x$ and $y$ are defined as the standard basis vectors in $\mathbb{R}^3$, along with $z$.), and by numbers such that sequentially numbered directions are orthogonal (modulo 5).}
\end{figure}
\end{center}

\noindent Represented by pairwise orthogonal directions (i.e. 1 is orthogonal to 2 and 5, etc.) which can be visualized as a pentagram, see Fig. \ref{pent1}.

\noindent If we consider the realist world view to be correct, we must consistently assign values to all these potential measurements. If we try to minimize this function we discover that there is a limit given by

\begin{eqnarray}
a_{1}a_{2}+a_{2}a_{3}+a_{3}a_{4}+a_{4}a_{5}+a_{5}a_{1}\geq -3.
\end{eqnarray}

\noindent To see why this must be the case, first assign values $a_{1}=+1$ and $a_{2}=-1$, minimizing the first term. Non-contextual realism then requires we also make the assignment $a_{2}=-1$ in the second term, so to minimize the second term we make the choice $a_{3}=+1$, and so on. Proceeding this way we discover that we are required to have at least one term be equal to $+1$, meaning that the series of measurements can not yield a result below negative three, likewise for the statistical averages

\begin{eqnarray}
\overline{a_{1}a_{2}}+\overline{a_{2}a_{3}}+\overline{a_{3}a_{4}}+\overline{a_{4}a_{5}}+\overline{a_{5}a_{1}}\geq -3.
\end{eqnarray}

\noindent To make an observation violating this inequality would be to exclude non-contextual realism as a valid world-view. Quantum mechanically this expression is state dependent, but if we choose a ``symmetric state" (that is a state represented as a vector aligned with the symmetry axis of the five directions) we find that, indeed, this inequality is violated

\begin{eqnarray}
\langle a_{1}a_{2}\rangle+\langle a_{2}a_{3}\rangle+\langle a_{3}a_{4}\rangle+\langle a_{4}a_{5}\rangle+\langle a_{5}a_{1}\rangle \simeq -3.944.
\end{eqnarray}

\noindent Where the triangle brackets represent quantum mechanical expectation values, in contrast with the over-bars which will be used to represent statistical averages.

\section{A Contextual Pentagram Inequality}

\noindent Non-contextuality states that, if we perform a specific measurement, along with a second co-measurable (commuting) measurement, then the second measurement cannot affect the first. With regard to the pentagram inequality, non-contextuality demands that we must assign the same value to $a_{x}$ in both $a_{x}a_{y}$ and $a_{x}a_{w}$. However it could be argued that though classical mechanics is non-contextual, some hypothetical hidden variable model is not. Thus if we relax the constraint of non-contextuality we must add a new label to each measurement of the form $a_{xy}a_{yx}$, such that the first index labels the measurement being performed and the second index labels the context. Now, the hidden variable model may assign values to all the elements of the series of measurements completely arbitrarily. The newly contextualized pentagram inequality becomes

\begin{eqnarray}
\overline{a_{12}a_{21}}+\overline{a_{23}a_{32}}+\overline{a_{34}a_{43}}+\overline{a_{45}a_{54}}+\overline{a_{51}a_{15}}\geq -5.
\end{eqnarray}

\noindent Which reaches below even the quantum limit. \\

\noindent However, in place of the restraint of non-contextuality we may add the requirement that the quantum mechanical spin-projection laws be obeyed. This is done in close analogy with the Leggett inequality, which deals with an entangled pair of spin-1/2 particles (photons in the polarization degrees of freedom) and relaxes locality (a form of contextuality), but requires that Malus' Law be obeyed by the individual particles. In the single-photon, three-rail analog of a spin-1 particle the rule that must be enforced is: The marginal probability for detection of a photon in a particular spatial mode \--- that is the probability that the photon be found in that mode regardless of other conditions \--- must obey the quantum mechanical projection rule. Physically this could be seen as the result of photons obeying the proper beam-splitter operations \--- something experimentally testable and understood classically. Mathematically $|\langle\psi|j\rangle|^{2}$, where $|\psi\rangle$ is the state vector and $|j\rangle$ is the optical mode being measured, visualized as a direction in $\mathbb{R}^{3}$. Though we use the language of QM, this could be formulated classically. This assumption, in a sense, is stronger than non-contextuality, since it is an assumption on the ``back end" as opposed to an assumption on the ``front end". That is, the constraint is one that deals with experimental results, and involves no assumptions about the fundamental nature of the theory. \\

\noindent We now derive an inequality which uses the spin projection assumption, but not the non-contextuality assumption. First we quickly derive a rule we will need. It involves the unused ``$z$" modes. Specifically, we require that if a photon is not found to be in either of the two observed modes, that it be in the unobserved mode. Or, mathematically, $P(^{-}_{j+1},^{-}_{j})=P(^{-}_{j+1})P(^{-}_{j}|^{-}_{j+1})=P(^{+}_{``z"})$. Where $P(^{-}_{j+1})$ is the probability that a measurement on $j+1^{\mathrm{th}}$ optical mode will yield a negative result (that is, not contain a photon). Likewise a $+$ represents the probability that the mode will contain a photon. We make use of the standard Bayesian notation for conditional probabilities. In order to ``break" the original Klyachko inequality, all that was needed was \emph{setting} dependence, however the expression we have written is \emph{outcome} dependent. The ``$z$" is in quotes because it stands for \emph{whichever} direction is mutually orthogonal to the two measurement directions in question. For symmetric states $P(^{+}_{``z"})$ is the same for each orthogonal pair (as the angle between all five vectors and the symmetric vector is the same), we will denote this number by the real constant $q$. Thus, we obtain

\begin{eqnarray}
P(^{-}_{j}|^{-}_{j+1})=\frac{P(^{+}_{``z"})}{P(^{-}_{j+1})}=\frac{q}{1-c}\label{r2}
\end{eqnarray}

\noindent This result utilizes the fact that the chance that a photon will be found in any particular mode is $c\equiv|\langle\psi|j\rangle|^{2}$ for symmetric states, and thus the chance that it will not be in that mode is $P(_{j}^{-})=1-c$ for all $j$'s.\\

\noindent Now we can begin with the derivation. Start with the contextualized series of five measurements, written as a sum

\begin{eqnarray}
\overline{a_{12}a_{21}}+\overline{a_{23}a_{32}}+\overline{a_{34}a_{43}}+\overline{a_{45}a_{54}}+\overline{a_{51}a_{15}}\nonumber\\
=\sum_{j=1}^{5}\overline{a_{j,j+1}a_{j+1,j}}.
\end{eqnarray}

\noindent Where again the sum is modulo five. Using the standard inequality two-outcome measurement outcomes $\overline{AB}\geq|\overline{A}+\overline{B}|-1$ we have

\begin{eqnarray}
\sum_{j=1}^{5}\overline{a_{j,j+1}a_{j+1,j}}\geq\sum_{j=1}^{5}\left|\overline{a_{j,j+1}}+\overline{a_{j+1,j}}\right|-5.
\end{eqnarray}

\noindent Now using a few successive applications of the triangle inequality $|a|+|b|\geq|a+b|$,

\begin{eqnarray}
\sum_{j=1}^{5}\overline{a_{j,j+1}a_{j+1,j}}\geq\left|\sum_{j=1}^{5}\left(\overline{a_{j,j+1}}+\overline{a_{j+1,j}}\right)\right|-5.
\end{eqnarray}

\noindent Each average may be rewritten in terms of the probabilities of the potential outcomes and the values of those outcomes (simply $+1$ and $-1$), as

\begin{eqnarray}
\overline{a_{j,j+1}}&=&P(^{-}_{j+1})P(^{+}_{j}|^{-}_{j+1})-P(^{-}_{j+1})P(^{-}_{j}|^{-}_{j+1})\nonumber\\
& &-P(^{+}_{j+1})P(^{-}_{j}|^{+}_{j+1}).
\end{eqnarray}

\noindent Therefore

\begin{widetext}
\begin{eqnarray}
\sum_{j=1}^{5}\overline{a_{j,j+1}a_{j+1,j}}&\geq&\left|\sum_{j=1}^{5}\left(P(^{-}_{j+1})P(^{+}_{j}|^{-}_{j+1})-P(^{-}_{j+1})P(^{-}_{j}|^{-}_{j+1})-P(^{+}_{j+1})P(^{-}_{j}|^{+}_{j+1})\right.\right.\nonumber\\
& &\left.\left.+P(^{-}_{j})P(^{+}_{j+1}|^{-}_{j})-P(^{-}_{j})P(^{-}_{j+1}|^{-}_{j})-P(^{+}_{j})P(^{-}_{j+1}|^{+}_{j})\right)\right|-5\nonumber\\
&\geq&\left|\sum_{j=1}^{5}\left(P(^{-}_{j+1})P(^{+}_{j}|^{-}_{j+1})-P(^{-}_{j+1})P(^{-}_{j}|^{-}_{j+1})-P(^{+}_{j+1})P(^{-}_{j}|^{+}_{j+1})\right.\right.\nonumber\\
& &\left.\left.+P(^{+}_{j+1})P(^{-}_{j}|^{+}_{j+1})-P(^{-}_{j+1})P(^{-}_{j}|^{-}_{j+1})-P(^{-}_{j+1})P(^{+}_{j}|^{-}_{j+1})\right)\right|-5\nonumber\\
&\geq& 2(1-c)\left|\sum_{j=1}^{5}P(^{-}_{j}|^{-}_{j+1})\right|-5.
\end{eqnarray}
\end{widetext}

\noindent Where, after the second inequality, we have used Bayes' rule for conditional probabilities

\begin{eqnarray}
P(A|B)=P(B|A)\frac{P(A)}{P(B)}.
\end{eqnarray}

\noindent In the second line we have simplified the expression and invoked the spin projection rules on the marginals. Now using the derived rule, Eq.(\ref{r2}), we obtain

\begin{eqnarray}
\sum_{j=1}^{5}\overline{a_{j,j+1}a_{j+1,j}}\geq10q-5\simeq-3.944.
\end{eqnarray}

\noindent Which is exactly the quantum mechanical result for symmetric states. This proves that a contextual hidden variable model for an analog single spin-1 particle \--- restricted by spin projection rules \--- is capable of reaching the quantum mechanical result. Thus such a model is not in contradiction with quantum mechanics, unlike an entangled spin-1/2 system.\\

\noindent It is worth pointing out a similarity with a derivation in the original KCBS paper \cite{pent}. The authors of that paper show that the pentagram inequality can be violated by a symmetric state undergoing a series of projections on the spin-zero case (of the three possible spin states of a spin-1 particle) of the form $|\langle\mathcal{L}|\psi\rangle|^{2}$, $|\mathcal{L}\rangle$ being the eigenstate to be projected on. The measurement in the KCBS case constitutes a projection onto the state of the entire system, whereas in our case the projection is only on a measurement event in a single optical mode. The two become equivalent if no distinction is made between marginal and joint probabilities \--- but this is precisely the case we consider. So, the inequality we derive here can be seen as a contextual-realistic replication of this violation in the extremal (equality fulfilling) case without recourse to use of the full quantum formalism of the spin-1 system.

\section{An Explicit Contextual Hidden Variable Model}

\noindent The previous section demonstrated that a contextual hidden variable model could, in principle, replicate the results of quantum mechanics for a single spin-1 particle and a series of five measurements. In this section we present one such explicit model, based on the non-local hidden variable model of Leggett. The model is not elegant or intuitive, but it has the advantage of being ``both ways" contextual. That is, each measurement is contextual on the other, and it is not necessary to causally order the events. A much simpler model could be presented (one which is almost identical to Leggett's), but it would not have the symmetry property of the one presented here. It is worthwhile to mention that it is not necessary to understand the mechanics of the model to follow the discussion that will come after. It is presented only for the benefit of the interested reader.\\

\noindent Roughly, what is necessary is that each unique ordering of setting and context vectors be mapped into two real ``threshold values" the relative sizes of which determine the necessary statistics.\\

\noindent For a measurement labeled by both a direction ($i$) and a context ($j$)

\begin{eqnarray}
&A_{ij}&\left(\lambda,\lambda_{ij}(\vec{i},\vec{j},\vec{\psi}),\gamma_{ij}(\vec{i},\vec{j},\vec{\psi})\right)\equiv\nonumber\\
& &\left\{
  \begin{array}{ll}
    -1 \quad\mathrm{for} & \lambda\in\left[\lambda_{ij},\gamma_{ij}\right]; \\
    +1 \quad\mathrm{for} & \lambda\in \left[0,\lambda_{ij}\right)\mathrm{or}\left(\gamma_{ij},1\right].
  \end{array}
\right.
\end{eqnarray}

\noindent Where $\vec{i}$, and $\vec{j}$ are the vectors representing the measurement direction, and the context, respectively; $\vec{\psi}$ is the vector representing the state. The parameter $\lambda$ is the hidden variable, we place no restrictions on its distribution other than it is real number bound between zero and one. The numerical values $\lambda_{ij}$, and $\gamma_{ij}$ are thresholds which determine what proportion of values for the hidden variable result in either an outcome of $-1$ or $+1$. The first threshold value is given by

\begin{eqnarray}
\gamma_{ij}=\lambda_{ij}-\left|\vec{i}\cdot\vec{\psi}\right|^{2}+1.
\end{eqnarray}

\noindent Without yet defining the second threshold value we can already see that the marginals yield the correct expression

\begin{eqnarray}
\overline{A_{ij}}=\int d\lambda A_{ij}&=&+1\int^{\lambda_{ij}}_{0}d\lambda-1\int^{\gamma_{ij}}_{\lambda_{ij}}d\lambda+1\int^{1}_{\gamma_{ij}}d\lambda\nonumber\\
&=&2\lambda_{ij}-2\gamma_{ij}+1=2\left|\vec{i}\cdot\vec{\psi}\right|^{2}-1.
\end{eqnarray}

\noindent The second threshold value is given by

\begin{widetext}
\begin{eqnarray}
\lambda_{ij}&=&H\left(\left|\vec{j}\cdot\vec{\psi}\right|^{2}-\left|\vec{i}\cdot\vec{\psi}\right|^{2}\right)\left|\vec{i}\cdot\vec{\psi}\right|^{2}+\delta\left(\left|\vec{j}\cdot\vec{\psi}\right|^{2}-\left|\vec{i}\cdot\vec{\psi}\right|^{2}\right)\nonumber\\
& &\times\left\{\left[\frac{\left|\vec{i}\cdot\vec{\psi}\right|^{2}}{2}\left(1+\frac{\left(\vec{j}\times\vec{i}\right)\cdot\vec{\psi}}{\left|\left(\vec{j}\times\vec{i}\right)\cdot\vec{\psi}\right|}\right)\right]+\delta\left(\left(\vec{j}\times\vec{i}\right)\cdot\vec{\psi}\right)\frac{1}{2}\left[1+\left(\hat{R}_{\vec{v}}\left(\frac{\pi}{2}\right)\left[\vec{j}\times\vec{i}\right]\right)\cdot\vec{\psi}\right]\left|\vec{i}\cdot{\psi}\right|^{2}\right\}
\end{eqnarray}
\end{widetext}

\noindent Where

\begin{eqnarray}
H(t)&\equiv&\left\{
         \begin{array}{ll}
           1, & \mathrm{for}\quad t>0; \\
           0, & \mathrm{for}\quad t\leq0,
         \end{array}
       \right. \\
\delta(t)&\equiv&\left\{
         \begin{array}{ll}
           1, & \mathrm{for}\quad t=0; \\
           0, & \mathrm{for}\quad t\neq0.
         \end{array}
       \right.
\end{eqnarray}

\noindent And $\hat{R}_{\vec{v}}(\theta)$ is a rotation operator which rotates a vector around the vector $\vec{v}$ (defined shortly) by angle $\theta$. Key to seeing how this model operates is that $A_{ij}$'s threshold values must be different from $A_{ji}$'s, otherwise attempting to integrate over the possible values of $\lambda$ for correlation values between the two will always yield zero \--- so the model must have some built-in asymmetry with regards to ordering; thus the step functions and cross products. To further understand this formulation, we should consider each term within the context of when it is non-zero. The $H$ \--- Heaviside step function \--- is only ``switched on" (i.e. non-zero) when the state projection onto the context vector is larger than the projection onto the measurement direction than on the actual measurement direction. For symmetric states (especially important for our analysis) \--- that is those with equal projections onto the context and direction vectors \--- the first delta function switches on. This delta function is distributed across two terms. The first term contains cross products which enforce the necessary asymmetry and yields the correct threshold values when the symmetrically projecting state vector is not in the plane defined by the $i$ and $j$ directions. However when it is in-plane this term is zero and the next term switches on. The final term contains a rotation operator in real-space $\hat{R}_{\vec{v}}$ which rotates about a vector which is perpendicular to the plane defined by the set of all possible symmetric states (again, here symmetric means only symmetric with regard to $\vec{i}$ and $\vec{j}$) by $\pi/2$ radians. This rotation allows the hidden variable model to assign working threshold values in a similar fashion to the previous term.\\

\noindent Now, if it is the case, for example, that $\left|\vec{j}\cdot\vec{\psi}\right|^{2}>\left|\vec{i}\cdot\vec{\psi}\right|^{2}$, then $\lambda_{ij}=\left|\vec{i}\cdot\vec{\psi}\right|^{2}$, $\lambda_{ji}=0$, $\gamma_{ij}=1$, and $\gamma_{ji}=1-\left|\vec{j}\cdot\vec{\psi}\right|^{2}$.\\

 \noindent Now we can compute the correlation of $A_{ij}$ and $A_{ji}$. For orthogonal measurement directions it must be the case that $\lambda_{ij}\leq\gamma_{ji}$, so we have

\begin{eqnarray}
\overline{A_{ij}A_{ji}}&=&-1\int^{\lambda_{ij}}_{0}d\lambda+1\int^{\gamma_{ji}}_{\lambda_{ij}}d\lambda-1\int^{1}_{\gamma_{ji}}d\lambda ,\nonumber\\
&=&-2(\lambda_{ij}-\gamma_{ji})-1,\nonumber\\
&=&-2\left(\left|\vec{j}\cdot\vec{\psi}\right|^{2}+\left|\vec{i}\cdot\vec{\psi}\right|^{2}\right)+1.
\end{eqnarray}

\noindent Which matches the quantum mechanical expression. If we had the opposite case, $\left|\vec{j}\cdot\vec{\psi}\right|^{2}<\left|\vec{i}\cdot\vec{\psi}\right|^{2}$, the model would have produced the same result. Now for the case of symmetric states \--- where the above formulation would break down \--- the step function is zero and the second term switches on. The factor $\left(\vec{j}\times\vec{i}\right)\cdot\vec{\psi}/\left|\left(\vec{j}\times\vec{i}\right)\cdot\vec{\psi}\right|$, produces a plus sign for one ordering of $i$ and $j$ and a negative for the other \--- creating the necessary asymmetry. The averaging procedure for the correlations then proceeds exactly as above. However for states that are both symmetric, and in the plane defined by both measurement directions, the previous formulation breaks down, and the second delta function switches on. The vector $\vec{i}\times\vec{j}$ is rotated such that it also is in the plane defined by $\vec{i}$ and $\vec{j}$ meaning that for one ordering of $i$ and $j$ in the threshold values the term factor is equal to one and for the other it is zero. Again, finding the average of the correlations proceeds as above and reproduces the correct quantum mechanical expression. Thus this hidden variable model \--- though perhaps overcomplicated \--- reproduces quantum mechanics under all possible circumstances involving a spin-1 system with two contexts, and is completely internally consistent. The addition of more measurement directions to the inequality can also be simulated by the HVM as all projectors have the same statistics as QM.\\

\section{Discussion and Conclusions}

\noindent The question now is why this procedure fails to find a contradiction with quantum physics. In the case of the entangled pair of spin-1/2 particles such a contradiction naturally arises. What is different about the spin-1 system?\\

\noindent The inequality derived in section three states that a contextual hidden variable model, which is constrained by the spin projection rules, \emph{may} reproduce the results of quantum mechanics and reach the floor of -3.994 for symmetric states in the pentagram inequality. Section four shows that such a hidden variable model does exist explicitly. It is significant that the inequality derived utilizes outcome dependent contextuality, whereas the explicit model only requires setting dependence to replicate quantum mechanics. Thus it could be inferred that a version of the contextual pentagram inequality could be derived which utilizes only setting dependence and still reaches the -3.944 floor. It is intriguing to note that the restriction to spin projection rules is powerful enough to bring the contextual floor up to -3.944 from -5.\\

\noindent Perhaps some insight may be gained by examining the recently articulated principle of ``global exclusive disjunction" \cite{ac}. Briefly, the principle of exclusive disjunction states that the sum of probabilities of events that are pair-wise exclusive can not be larger than one. \emph{Global} exclusive disjunction states that this principle must be upheld when events in an inequality are considered jointly with other events, and places a lower bound on the quantum-contextual pentagram inequality. For more details see the cited reference. Since the explicit hidden variable model outlined in this manuscript replicates the quantum probabilities for both marginals and conditionals it follows that exclusive disjunction (and consequently global exclusive disjunction) is satisfied. This can be seen as leading to the lower bound on the explicit model.\\

\noindent Another point of interest is that, since this work drew inspiration from the Leggett inequality for entangled spin-1/2 particles, it would make sense that \--- in similarity with Leggett \--- some non-standard rotation of the the state vector in $\mathbb{C}^{3}$ would yield a contradiction with quantum mechanics (in the case of Leggett the polarization measurement projection vector must be rotated outside of the real plane of the Poincar\'{e} sphere to achieve violation). This is in fact not the case as the explicit model can recover the quantum mechanical results for \emph{any} state vector in a two-context system. It is unknown why this is, it may be that the entanglement of two subsystems is more ``powerful" than the spin-1 particle. Or there may be some altogether different cause. Perhaps a generalization to more contexts will yield a contradiction with quantum theory. These questions remain open and we hope this manuscript stimulates further interest in this issue.\\

\section*{Acknowledgements}

\noindent The authors would like to acknowledge Johannes Kofler for many useful discussions, as well as Anton Zeilinger for further discussions and support. This work was supported by the ERC Advanced Grant QIT4QAD, and the Austrian Science Fund FWF within the SFB F40 (FoQuS) and W1210-2 (CoQuS).

\end{document}